\documentclass[sigconf,nonacm,natbib=true,pbalance]{acmart}
\AtBeginDocument{%
  }

\usepackage{csquotes} 
\usepackage{pbalance}

\newsavebox{\promptbox}
\newenvironment{prompt}{%
  \par\smallskip\noindent
  \begin{lrbox}{\promptbox}%
  \begin{minipage}{\dimexpr\linewidth-2\fboxsep-2\fboxrule\relax}%
  \setlength{\parindent}{0pt}%
  \ignorespaces
}{%
  \end{minipage}%
  \end{lrbox}%
  \setlength{\fboxsep}{8pt}%
  \setlength{\fboxrule}{0.4pt}%
  \fcolorbox{gray!50}{gray!8}{\usebox{\promptbox}}%
  \par\smallskip
}

\makeatletter
\newcommand{\acmrightssize}{\fontsize{8}{9.5}\selectfont}

\newcommand{\firstpagerights}[1]{%
  \begingroup
    \renewcommand\thefootnote{}%
    \footnotetext{%
      \acmrightssize
      \raggedright
      \setlength{\parskip}{0pt}%
      \setlength{\parindent}{0pt}%
      #1%
    }%
    \addtocounter{footnote}{0}%
  \endgroup
}
\makeatother

\begin{document}

\title{In-Browser Agents for Search Assistance}

\author{Saber Zerhoudi}
\orcid{0000-0003-2259-0462}
\affiliation{%
  \institution{University of Passau}
  \city{Passau}
  \country{Germany}
}
\email{saber.zerhoudi@uni-passau.de}

\author{Michael Granitzer}
\orcid{0000-0003-3566-5507}
\affiliation{%
  \institution{University of Passau}
  \city{Passau}
  \country{Germany}
}
\affiliation{%
  \institution{Interdisciplinary Transformation University Austria}
  \city{Linz}
  \country{Austria}
}
\email{michael.granitzer@uni-passau.de}

\renewcommand{\shortauthors}{S. Zerhoudi et al.}

\begin{abstract}
A fundamental tension exists between the demand for sophisticated AI assistance in web search and the need for user data privacy. Current centralized models require users to transmit sensitive browsing data to external services, which limits user control. In this paper, we present a browser extension~\footnote{\url{https://github.com/saberzerhoudi/agentic-search-plugin}} that provides a viable in-browser alternative. We introduce a hybrid architecture that functions entirely on the client side, combining two components: (1) an adaptive probabilistic model that learns a user's behavioral policy from direct feedback, and (2) a Small Language Model (SLM), running in the browser, which is grounded by the probabilistic model to generate context-aware suggestions. To evaluate this approach, we conducted a three-week longitudinal user study with 18 participants. Our results show that this privacy-preserving approach is highly effective at adapting to individual user behavior, leading to measurably improved search efficiency. This work demonstrates that sophisticated AI assistance is achievable without compromising user privacy or data control.
\end{abstract}

% --- ACM CCS Concepts (put after the abstract, before \maketitle) ---
% If your venue requires CCSXML, we can add it, but \ccsdesc lines are enough to compile.


\begin{teaserfigure}
  \begin{center}
\includegraphics[width=0.9\textwidth]{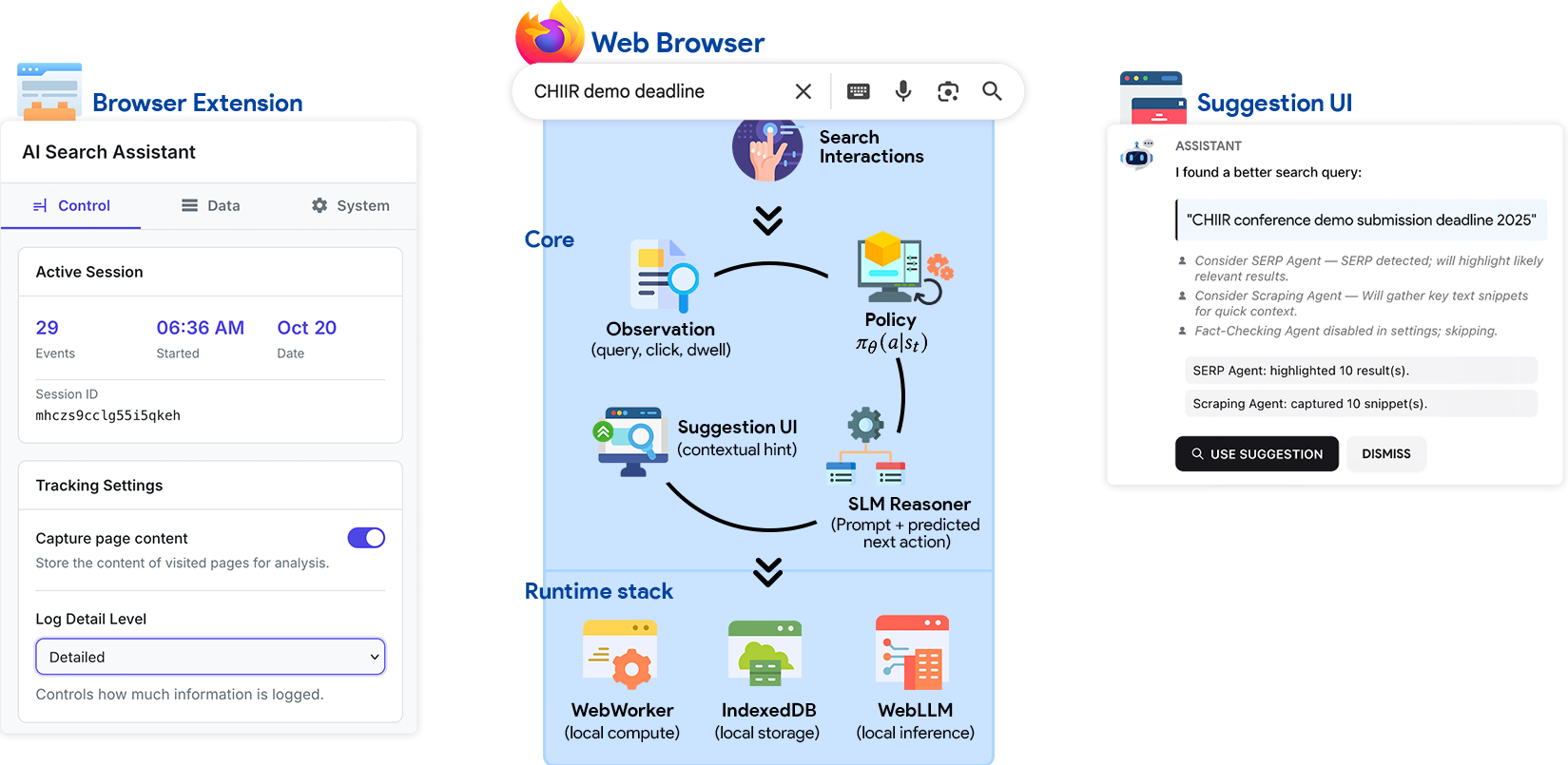}
\vspace{-3mm}
  \caption{Overview and Interfaces of the In-Browser, Behavior-Grounded Search Assistant.}
  \label{fig_simiir3}
  \vspace{4mm}
\end{center}
\end{teaserfigure}

\maketitle
\firstpagerights{%
  © ACM, 2026. This is the author's version of the work.\\
  The definitive version was published in:
  \emph{Proceedings of the 2026 ACM SIGIR Conference on Human Information Interaction and Retrieval (CHIIR '26),
  March 22--26, 2026, Seattle, WA, USA}.\\
  DOI: \url{https://doi.org/10.1145/3786304.3787913}
}

\section{Introduction}

The incorporation of agentic AI within web browsers is altering information-seeking behaviors~\cite{Microsoft:AI-Powered-Bing-Edge2023,Google:Generative-AI-in-Search2024,OpenAI:ChatGPT-Atlas2025}; however, this development is solidifying a centralized, cloud-based infrastructure. Such an operational model depends on transmitting a continuous stream of sensitive user data---including browsing history, queries, and interactions---for external processing~\cite{NarayananS08:SP,ohm2009broken}. Consequently, users are forced into a difficult position, having to select either sophisticated AI assistance or the control of their own data. In addition, using massive, general-purpose Large Language Models (LLMs) represents an inefficient use of computational resources for the specialized tasks in search assistance~\cite{vannguyen2024surveysmalllanguagemodels,belcak2025smalllanguagemodelsfuture}. This creates two central problems: first, a lack of privacy and user control, and second, an incorrect match between massive, high-cost models and the precise, contextual functions they are meant to perform.

In this work, we contend that recent technological developments offer a solution to this conflict. This solution is supported by two developments: first, the emergence of efficient, specialized Small Language Models (SLMs) suited for specific domains~\cite{abdin2024phi3technicalreporthighly,allal2025smollm2smolgoesbig,belcak2025smalllanguagemodelsfuture}, and second, the new-found capability to run these SLMs directly in the browser using technologies such as WebGPU~\cite{ruan2024webllmhighperformanceinbrowserllm,Chen0SL25:WWW}. By leveraging these two trends, we introduce a new framework that provides adaptive search assistance \emph{while ensuring no data exits the user's local device}. Our proposed solution is a hybrid, fully client-side architecture. Within this structure, a predictive probabilistic model learns the user's policy, and its outputs are then used as a grounding mechanism for an SLM reasoner that operates within the browser~\cite{ruan2024webllmhighperformanceinbrowserllm}.

This paper provides the following contributions:
\begin{itemize}
    \item We design and implement a novel, fully in-browser architecture that grounds a local SLM reasoner with an adaptive, predictive probabilistic model.
    \item We introduce an in-browser online learning mechanism~\cite{Cacciarelli_2023,lu2025webservbrowserserverenvironmentefficient} that effectively personalizes a generic user model to an individual's specific search patterns using direct feedback.
    \item We demonstrate through a 3-week longitudinal study that this fully client-side framework measurably improves search efficiency without compromising user privacy or agency.
    \item We release our system as an open-source tool to facilitate further research in privacy-centric, in-browser agentic AI.
\end{itemize}

\section{Related Work}
Our work is positioned at the intersection of probabilistic user modeling, agent architectures, and in-browser language models.

\subsection{Probabilistic User Models}
Modeling search behavior using methods like click models and Markov models has proven effective for capturing aggregate user patterns~\cite{ChierichettiKRS12, awad2012prediction, tran2013markov, ZerhoudiGSS22}. However, these models are traditionally predictive, not generative, and often lack mechanisms for real-time adaptation to an individual's intent~\cite{farshidi2023understandinguserintentmodeling}. Our work extends this approach: we use a lightweight, adaptive probabilistic model as the \emph{grounding mechanism} for a generative reasoning engine~\cite{ZhangWGLM24}.

\subsection{LLM-Driven Information Seeking}
The application of LLMs to search tasks is promising~\cite{becker2024multiagentlargelanguagemodels,ZhangWGLM24}, but current methods rely on cloud-based APIs. This architecture re-introduces privacy risks~\cite{NarayananS08,ohm2009broken}, and can produce ungrounded outputs detached from the user's immediate context~\cite{Ji_2023}. Our approach is distinct in that it \emph{rejects} this cloud-dependent model by bringing the reasoning engine to the client, solving the privacy, and grounding problems simultaneously.

\subsection{In-Browser and Small Language Models}
Our work is made possible by two concurrent trends. First, the shift toward highly efficient, specialized Small Language Models (SLMs)~\cite{vannguyen2024surveysmalllanguagemodels} as a sustainable and effective solution for vertical domains like search assistance~\cite{abdin2024phi3technicalreporthighly}. Second, advancements in web technologies (e.g., WebGPU~\cite{kenwright2022introduction,Chen0SL25:WWW}, WebLLM~\cite{ruan2024webllmhighperformanceinbrowserllm}) now permit these multi-billion parameter SLMs to be executed directly within the browser. While some in-browser models exist for recommendation~\cite{wang2024anatomizingdeeplearninginference,Zhao_2024}, they typically lack generative reasoning. To our knowledge, no prior work has unified an adaptive behavioral model with a local SLM reasoner in a fully client-side architecture for search assistance. Our framework is designed to fill this gap.

\section{System Architecture and Methodology}
We designed our framework to solve the challenge of creating a grounded, adaptive, and fully private search assistant. The entire architecture is implemented as a Firefox browser extension. The fundamental design principle of this system is to perform all computation---from data-logging to probabilistic modeling and generative inference---entirely on the client side. This method ensures that no sensitive user data leaves the user's machine, thereby directly resolving the conflict between personalization and privacy.

The architecture combines the predictive accuracy of a lightweight probabilistic model with the generative reasoning capabilities of an in-browser Small Language Model (SLM). It operates as a passive assistant, observing user behavior and producing contextual suggestions. These suggestions are based on a dynamic, locally-stored understanding of the user's state and evolving goals. The system is composed of three primary modules that work together:
\begin{itemize}
    \item Behavioral Observation and Provenance.
    \item Dynamic User Modeling and State Estimation.
    \item Cognitive Inference and Suggestion.
\end{itemize}

\subsection{Behavioral Observation and Provenance}
The framework's foundation is a perceptual layer that captures a user's interaction data within an active search session. This module operates as a client-side listener, logging a curated set of high-level interaction events indicative of an information-seeking strategy:
\begin{itemize}
    \item \textbf{Issued Queries:} The raw text of each query.
    \item \textbf{SERP Clicks:} The rank and URL of each click on a Search Engine Result Page (SERP).
    \item \textbf{Document Dwell Times:} The time spent on a document before returning to the SERP.
\end{itemize}
This data is stored locally using the browser's \texttt{IndexedDB} API~\cite{W3C_IndexedDB_2015}, a persistent, in-browser database. This design choice is central to our privacy-preserving commitment. To ensure complete user agency, the extension's interface provides clear controls for the user to enable, disable, or clear all logged data. The user retains sole authority over their data.

\subsection{Dynamic User Model}
This module is the analytical core of the framework. It transforms raw interaction logs into an adaptive policy model that can predict the user's subsequent actions. We formalize the search session as a \textbf{Markov Decision Process (MDP)}, which provides a robust mathematical framework for modeling sequential decision-making.

\paragraph{States and Actions} A session is defined by states $s_t \in S$ and actions $a_t \in A$. The state $s_t$ represents the user's current context (e.g., \texttt{\small ViewingSERP}, \texttt{\small ReadingDocument}). The action space $A$ is discrete and includes strategic actions such as \texttt{\small ClickDocument\_i} (for $i \in [1, 10]$), \texttt{\small ReturnToSERP}, and \texttt{\small SubmitNewQuery(}\textit{\small type}\texttt{\small)}, where \textit{type} is classified as \textit{generalization}, \textit{specialization}, or \textit{reformulation}.

\paragraph{Model Initialization (Cold Start)} To address the cold-start problem and ensure the system provides immediate value, it is provisioned with pre-trained behavioral policies, $\pi_{\theta}(a|s)$. These policies function as representations of established searcher archetypes, such as ``exploratory'' or ``lookup''~\cite{agarwal2021lookupexploratorysearchintent}. We generated these initial policies by training a compact \textbf{Multi-Layer Perceptron (MLP)}~\cite{li2024bmlpbehaviorawaremlpheterogeneous} on clustered sessions from the AOL query log dataset~\cite{AOLQueryLog2006}. The behavior-aware MLP was selected as the model architecture due to its minimal resource footprint and rapid inference capabilities, which makes it highly suitable for execution within the browser.

\paragraph{Prediction} The module's primary function is to use the current policy $\pi_{\theta}$ to compute a probability distribution over the next set of actions given the user's current state $s_t$. The action with the highest probability, $a^* = \arg\max_{a \in A} \pi_{\theta}(a|s_t)$, is identified as the most probable next action. This prediction is the critical output used to ground the SLM.

A crucial component of this module is its capacity for in-browser online learning via direct user feedback. This mechanism allows the model to evolve from a generic archetype into a policy that is deeply personalized. When a user accepts ($f_t = +1$) or discards ($f_t = -1$) a suggestion, the policy network's parameters $\theta$ are updated using a simple and efficient policy gradient rule:
\begin{equation}
\label{eq:pg_update}
\theta_{t+1} = \theta_t + \alpha \cdot f_t \cdot \nabla_{\theta} \log \pi_{\theta}(a_t|s_t)
\end{equation}
Here, $\alpha$ is the learning rate (set to $0.01$). This update follows the REINFORCE algorithm~\cite{Williams1992}, directly adjusting the likelihood of the action that produced the suggestion by either strengthening or weakening it, depending on the user's explicit feedback. The entire process, from prediction to update, runs within a dedicated \textbf{Web Worker}~\cite{W3C_WebWorkers_20210128}. This ensures that the model's computational tasks do not block the browser's main thread, thereby preventing any degradation of the user experience. The resulting updated policy parameters, $\theta_{t+1}$, are then saved locally in \texttt{IndexedDB}, which establishes a virtuous cycle of continuous personalization.

\subsection{Hybrid Cognitive Inference}
This module connects the probabilistic model's abstract prediction to the generative, human-readable advice the user receives. The predicted next action, $a^*$, is \emph{not} shown directly to the user. Instead, it serves as a \textit{grounding directive} for the in-browser SLM.

This two-step process is a key design choice. Instead of prompting an unconstrained SLM with a raw history to ``help the user'', our method first utilizes the probabilistic model to identify a high-likelihood \emph{strategic} action, $a^*$. It is this action, $a^*$, that then guides the SLM's reasoning process. This technique constrains the solution space, ensuring the generated advice is directly relevant to the user's predicted behavior.

The predicted action is inserted into a prompt template along with recent session history. A simplified prompt structure is:

\begin{prompt}
\small\ttfamily
You are a helpful search assistant. Based on the user's recent activity, 
they will likely perform the following action: 
\textbf{\{Predicted Action $a^*$\}}. 

The user's last query was ``\textbf{\{Last Query\}}''. 

Generate a concise, helpful suggestion to guide them.
\end{prompt}

The entire inference pipeline runs locally. We leverage the \textbf{WebGPU engine} to execute a \textasciitilde2.7B parameter model from the Phi family~\cite{abdin2024phi3technicalreporthighly} directly in the browser. This model was selected as it provides state-of-the-art balanced high performance with a suitable footprint for client-side execution~\cite{LuL0YLLLZLX25}. Our tests confirmed the viability of this approach on modern hardware~\footnote{Apple MacBook Pro (M1 Pro, 10-core CPU / 16-core GPU, 32 GB unified memory).}. The local SLM generated suggestions with an average latency of \textasciitilde9 tokens/sec, which is well within the threshold for a non-disruptive user experience.

Finally, to avoid ``suggestion fatigue'', the system incorporates an \textit{intrusion avoidance} heuristic. It refrains from generating suggestions when (1) the model predicts a clear navigational action with high confidence, or (2) the probability distribution over next actions is nearly uniform, indicating high model uncertainty.

\section{Experimental Evaluation}
Our evaluation was designed as a longitudinal study to validate the framework's core claims. The goals were to (1) quantitatively measure the effectiveness of the in-browser personalization mechanism and (2) assess the framework's real-world impact on both the search behavior and subjective experience of users.

\subsection{Experimental Setup}
\paragraph{Participants} We recruited 18 participants (12 male, 6 female, ages 21-32) from a local university, representing a range of technical backgrounds with high self-reported familiarity with web search. 

\paragraph{Procedure} We conducted a 3-week longitudinal study, which was critical for observing the model's adaptation over time.
\begin{itemize}
    \item \textbf{Week 1 (Baseline):} Participants installed the extension with suggestions \emph{disabled}. This allowed us to collect baseline data on their natural, un-assisted search behavior.
    \item \textbf{Weeks 2-3 (Intervention):} The suggestion feature was \emph{enabled}. Participants continued their normal web tasks, interacting with the suggestions as they chose.
\end{itemize}

\paragraph{Data Collection}
To ensure a realistic evaluation environment, participants used their own laptops running Firefox. The extension logged all anonymized interaction data locally. At the study's conclusion, participants \emph{themselves} exported and submitted this anonymized data. They also completed a post-study questionnaire assessing usability (SUS)~\cite{Brooke1996SUS}, perceived utility, and trust.

\subsection{Model Personalization and Adaptation}
\paragraph{Motivation} Our first research question was to determine if the in-browser online learning mechanism could effectively adapt a generic model to an individual's unique search patterns. We hypothesized that a personalized model would outperform the initial generic models in predicting that user's \emph{own} future actions.

\paragraph{Setting} We analyzed the interaction logs from the intervention phase using a temporal hold-out strategy: for each participant, the first 80\% of their interactions were used to simulate the adaptation process (training via Eq. \ref{eq:pg_update}), and the final 20\% served as the test set for a next-action prediction task.

\paragraph{Baselines} We compared the predictive performance of the pre-trained generic models (Generic-Exploratory, Generic-Lookup) against the personalized \textbf{Persona-Adapted} model for each user.

\paragraph{Results} The results, summarized in Table~\ref{tab:model_perf}, provide strong quantitative support for our hypothesis. The Persona-Adapted model (38.7\% Accuracy, 0.47 MRR) outperformed both generic baselines. Compared to the strongest baseline (Generic-Exploratory), our adapted model achieved a 24.0\% relative improvement in prediction accuracy and 20.5\% in MRR. This confirms that the online learning mechanism is successfully capturing user-specific search habits, evolving from a generic archetype into a specialized policy. 

\begin{table}[t]
\caption{Next-action prediction performance. The personalized in-browser model outperforms the static, generic baselines.}
\label{tab:model_perf}
\begin{tabular}{@{}lcc@{}}
\toprule
\textbf{Model} & \textbf{Pred. Acc. (\%)} & \textbf{MRR} \\
\midrule
Generic-Exploratory & 31.2 & 0.39 \\
Generic-Lookup & 26.5 & 0.34 \\
\textbf{Persona-Adapted} & \textbf{38.7} & \textbf{0.47} \\
\bottomrule
\end{tabular}
\end{table}

\subsection{Behavioral Shift and User Perception}
\paragraph{Motivation} Our ultimate goal is not just to predict actions, but to \emph{improve} the user's search process. We evaluated whether exposure to the assistant led to measurable changes in search behavior and assessed the user's subjective experience.

\paragraph{Setting} We compared behavioral metrics from the one-week Baseline phase (no suggestions) against the two-week Intervention phase using paired $t$-tests. We also analyzed the subjective ratings from the post-study questionnaire.

\paragraph{Results}
The analysis in Table \ref{tab:user_perf} revealed a statistically positive improvement in user search behavior. Participants' queries became measurably more complex during the intervention (4.1 vs. 3.5 terms, $p<0.05$), suggesting the assistant guided them to more descriptive formulations. This, in turn, led to a decrease in average session length (5.2 vs. 6.8 queries, $p<0.05$), indicating a clear gain in search efficiency.

The mean suggestion acceptance rate was 36.4\%. This result is notable as it shows users maintained their agency, evaluating suggestions rather than passively accepting them. Subjective feedback was highly positive, with an ``Excellent'' SUS score of 82.5. Critically for a privacy-focused tool, users reported both high perceived utility (6.1/7) and trust (5.9/7).

\begin{table}[t]
\caption{Comparison of behavioral and subjective metrics between the Baseline and Intervention phases. Asterisks (*) denote a statistically significant difference ($p<0.05$).}
\label{tab:user_perf}
\begin{tabular}{@{}lcc@{}}
\toprule
\textbf{Metric} & \textbf{Baseline} & \textbf{Intervention} \\
\midrule
Avg. Query Complexity (terms) & 3.5 & 4.1* \\
Avg. Session Length (queries) & 6.8 & 5.2* \\
Suggestion Acceptance Rate (\%) & - & 36.4 \\
System Usability Scale (SUS) & - & 82.5 \\
Perceived Utility (1--7) & - & 6.1 \\
Trust in Suggestions (1--7) & - & 5.9 \\
\bottomrule
\end{tabular}
\end{table}

\section{Discussion and Conclusion}
Our experimental evaluation confirms the framework's effectiveness. The in-browser online learning mechanism successfully adapted the generic policy, yielding a 24.0\% improvement in next-action prediction accuracy over the strongest baseline. This personalization also measurably improved search efficiency: we observed a statistically reduction in average session length (from 6.8 to 5.2 queries) as users were guided toward more effective, complex search strategies.

A 36.4\% suggestion acceptance rate further validates our core architectural thesis. It demonstrates that our hybrid approach---combining a probabilistic model's behavioral predictions with an in-browser SLM's reasoning---provides useful, context-aware guidance while preserving complete user control.

Most importantly, this work challenges the prevailing assumption that sophisticated agentic assistance must be centralized. Our findings demonstrate a viable, computationally efficient, and fully private alternative. We show that by using specialized SLMs with in-browser execution, it is possible to achieve effective personalization without compromising user data sovereignty.

\subsection{Limitations and Future Work}
We acknowledge this study's limitations. The participant sample was drawn from a university population and is not representative of the general public. Furthermore, our evaluation was designed to isolate the effect of personalization, comparing the adapted model only against its own generic baseline rather than external methods.

Future work will focus on benchmarking our adaptive model against more complex sequential recommendation baselines to situate its predictive performance. We also plan to extend the framework to process multimodal content and to integrate more advanced reinforcement learning algorithms for policy adaptation, all while maintaining the fully client-side, privacy-preserving architecture.

\subsection{Conclusion}

This paper presented a novel framework for a privacy-preserving search assistant that operates entirely on the user's device. We designed a system that learns from and adapts to the user by grounding an in-browser SLM with an adaptive policy model. Our results show a viable method for providing sophisticated AI assistance while maintaining complete user autonomy. We provide our framework as an open-source tool to support further research and development in this user-centric, in-browser paradigm.

% ---------- Bibliography ----------
\bibliographystyle{ACM-Reference-Format}
\bibliography{main}

@String{Computing = "Computing" }

@String{Computer = "{IEEE} Computer" }

@String{Springer = "Springer-Verlag" }

@online{Microsoft:AI-Powered-Bing-Edge2023,
  author       = {Yusuf Mehdi},
  title        = {Reinventing search with a new {AI}-powered {Microsoft} Bing and Edge, Your Copilot for the Web},
  year         = {2023},
  month        = {Feb},
  day          = {7},
  url          = {https://blogs.microsoft.com/blog/2023/02/07/reinventing-search-with-a-new-ai-powered-microsoft-bing-and-edge-your-copilot-for-the-web/},
  note         = {Microsoft Corporation Blog}
}

@online{Google:Generative-AI-in-Search2024,
  author       = {Google Inc.},
  title        = {Generative {AI} in Search: Let {Google} do the searching for you},
  year         = {2024},
  month        = {May},
  day          = {20},
  url          = {https://blog.google/products/search/generative-ai-google-search-may-2024/},
  note         = {Google Blog}
}

@online{OpenAI:ChatGPT-Atlas2025,
  author       = {OpenAI},
  title        = {Introducing ChatGPT Atlas},
  year         = {2025},
  month        = {Oct},
  day          = {21},
  url          = {https://openai.com/index/introducing-chatgpt-atlas/},
  note         = {OpenAI Blog}
}

@inproceedings{NarayananS08:SP,
  author       = {Arvind Narayanan and
                  Vitaly Shmatikov},
  title        = {Robust De-anonymization of Large Sparse Datasets},
  booktitle    = {2008 {IEEE} Symposium on Security and Privacy {(SP} 2008), 18-21 May
                  2008, Oakland, California, {USA}},
  pages        = {111--125},
  publisher    = {{IEEE} Computer Society},
  year         = {2008},
  url          = {https://doi.org/10.1109/SP.2008.33},
  doi          = {10.1109/SP.2008.33}
}

@article{ohm2009broken,
  title={Broken promises of privacy: Responding to the surprising failure of anonymization},
  author={Ohm, Paul},
  journal={UCLA l. Rev.},
  volume={57},
  pages={1701},
  year={2009},
  publisher={HeinOnline}
}

@misc{vannguyen2024surveysmalllanguagemodels,
      title={A Survey of Small Language Models}, 
      author={Chien Van Nguyen and Xuan Shen and Ryan Aponte and Yu Xia and Samyadeep Basu and Zhengmian Hu and Jian Chen and Mihir Parmar and Sasidhar Kunapuli and Joe Barrow and Junda Wu and Ashish Singh and Yu Wang and Jiuxiang Gu and Franck Dernoncourt and Nesreen K. Ahmed and Nedim Lipka and Ruiyi Zhang and Xiang Chen and Tong Yu and Sungchul Kim and Hanieh Deilamsalehy and Namyong Park and Mike Rimer and Zhehao Zhang and Huanrui Yang and Ryan A. Rossi and Thien Huu Nguyen},
      year={2024},
      eprint={2410.20011},
      archivePrefix={arXiv},
      primaryClass={cs.CL},
      url={https://arxiv.org/abs/2410.20011}, 
}

@misc{belcak2025smalllanguagemodelsfuture,
      title={Small Language Models are the Future of Agentic AI}, 
      author={Peter Belcak and Greg Heinrich and Shizhe Diao and Yonggan Fu and Xin Dong and Saurav Muralidharan and Yingyan Celine Lin and Pavlo Molchanov},
      year={2025},
      eprint={2506.02153},
      archivePrefix={arXiv},
      primaryClass={cs.AI},
      url={https://arxiv.org/abs/2506.02153}, 
}

@misc{abdin2024phi3technicalreporthighly,
      title={Phi-3 Technical Report: A Highly Capable Language Model Locally on Your Phone}, 
      author={Marah Abdin and Jyoti Aneja and Hany Awadalla and Ahmed Awadallah and Ammar Ahmad Awan and Nguyen Bach and Amit Bahree and Arash Bakhtiari and Jianmin Bao and Harkirat Behl and Alon Benhaim and Misha Bilenko and Johan Bjorck and Sébastien Bubeck and Martin Cai and Qin Cai and Vishrav Chaudhary and Dong Chen and Dongdong Chen and Weizhu Chen and Yen-Chun Chen and Yi-Ling Chen and Hao Cheng and Parul Chopra and Xiyang Dai and Matthew Dixon and Ronen Eldan et al.},
      year={2024},
      eprint={2404.14219},
      archivePrefix={arXiv},
      primaryClass={cs.CL},
      url={https://arxiv.org/abs/2404.14219}, 
}

@misc{allal2025smollm2smolgoesbig,
      title={SmolLM2: When Smol Goes Big -- Data-Centric Training of a Small Language Model}, 
      author={Loubna Ben Allal and Anton Lozhkov and Elie Bakouch and Gabriel Martín Blázquez and Guilherme Penedo and Lewis Tunstall and Andrés Marafioti and Hynek Kydlíček and Agustín Piqueres Lajarín and Vaibhav Srivastav and Joshua Lochner and Caleb Fahlgren and Xuan-Son Nguyen and Clémentine Fourrier and Ben Burtenshaw and Hugo Larcher and Haojun Zhao and Cyril Zakka and Mathieu Morlon and Colin Raffel and Leandro von Werra and Thomas Wolf},
      year={2025},
      eprint={2502.02737},
      archivePrefix={arXiv},
      primaryClass={cs.CL},
      url={https://arxiv.org/abs/2502.02737}, 
}

@misc{ruan2024webllmhighperformanceinbrowserllm,
      title={WebLLM: A High-Performance In-Browser LLM Inference Engine}, 
      author={Charlie F. Ruan and Yucheng Qin and Xun Zhou and Ruihang Lai and Hongyi Jin and Yixin Dong and Bohan Hou and Meng-Shiun Yu and Yiyan Zhai and Sudeep Agarwal and Hangrui Cao and Siyuan Feng and Tianqi Chen},
      year={2024},
      eprint={2412.15803},
      archivePrefix={arXiv},
      primaryClass={cs.LG},
      url={https://arxiv.org/abs/2412.15803}, 
}

@inproceedings{Chen0SL25:WWW,
  author       = {Zhiyang Chen and
                  Yun Ma and
                  Haiyang Shen and
                  Mugeng Liu},
  editor       = {Guodong Long and
                  Michale Blumestein and
                  Yi Chang and
                  Liane Lewin{-}Eytan and
                  Zi Helen Huang and
                  Elad Yom{-}Tov},
  title        = {WeInfer: Unleashing the Power of WebGPU on {LLM} Inference in Web
                  Browsers},
  booktitle    = {Proceedings of the {ACM} on Web Conference 2025, {WWW} 2025, Sydney,
                  NSW, Australia, 28 April 2025- 2 May 2025},
  pages        = {4264--4273},
  publisher    = {{ACM}},
  year         = {2025},
  url          = {https://doi.org/10.1145/3696410.3714553}
}

@misc{lu2025webservbrowserserverenvironmentefficient,
      title={WEBSERV: A Browser-Server Environment for Efficient Training of Reinforcement Learning-based Web Agents at Scale}, 
      author={Yuxuan Lu and Jing Huang and Hui Liu and Jiri Gesi and Yan Han and Shihan Fu and Tianqi Zheng and Dakuo Wang},
      year={2025},
      eprint={2510.16252},
      archivePrefix={arXiv},
      primaryClass={cs.LG},
      url={https://arxiv.org/abs/2510.16252}, 
}

@article{Cacciarelli_2023,
   title={Active learning for data streams: a survey},
   volume={113},
   ISSN={1573-0565},
   url={http://dx.doi.org/10.1007/s10994-023-06454-2},
   DOI={10.1007/s10994-023-06454-2},
   number={1},
   journal={Machine Learning},
   publisher={Springer Science and Business Media LLC},
   author={Cacciarelli, Davide and Kulahci, Murat},
   year={2023},
   month=nov, pages={185–239} 
}

@inproceedings{ChierichettiKRS12,
  author       = {Flavio Chierichetti and
                  Ravi Kumar and
                  Prabhakar Raghavan and
                  Tam{\'{a}}s Sarl{\'{o}}s},
  editor       = {Alain Mille and
                  Fabien Gandon and
                  Jacques Misselis and
                  Michael Rabinovich and
                  Steffen Staab},
  title        = {Are web users really Markovian?},
  booktitle    = {Proceedings of the 21st World Wide Web Conference 2012, {WWW} 2012,
                  Lyon, France, April 16-20, 2012},
  pages        = {609--618},
  publisher    = {{ACM}},
  year         = {2012},
  url          = {https://doi.org/10.1145/2187836.2187919},
  doi          = {10.1145/2187836.2187919}
}

@inproceedings{tran2013markov,
  title={Markov modeling for user interaction in retrieval},
  author={Tran, Vu T and Fuhr, Norbert},
  booktitle={SIGIR 2013 Workshop on Modeling User Behavior for Information Retrieval Evaluation (MUBE 2013)},
  volume={5},
  number={3},
  pages={7--2},
  year={2013},
  organization={Citeseer}
}

@article{awad2012prediction,
  title={Prediction of user's web-browsing behavior: Application of markov model},
  author={Awad, Mamoun A and Khalil, Issa},
  journal={IEEE Transactions on Systems, Man, and Cybernetics, Part B (Cybernetics)},
  volume={42},
  number={4},
  pages={1131--1142},
  year={2012},
  publisher={IEEE}
}

@inproceedings{ZerhoudiGSS22,
  author       = {Saber Zerhoudi and
                  Michael Granitzer and
                  Christin Seifert and
                  J{\"{o}}rg Schl{\"{o}}tterer},
  editor       = {Giorgio Maria Di Nunzio and
                  Beatrice Portelli and
                  Domenico Redavid and
                  Gianmaria Silvello},
  title        = {Simulating User Interaction and Search Behaviour in Digital Libraries},
  booktitle    = {Proceedings of the 18th Italian Research Conference on Digital Libraries,
                  Padua, Italy, February 24-25, 2022 (hybrid event)},
  series       = {{CEUR} Workshop Proceedings},
  volume       = {3160},
  publisher    = {CEUR-WS.org},
  year         = {2022},
  url          = {https://ceur-ws.org/Vol-3160/paper8.pdf}
}

@incollection{kenwright2022introduction,
  title={Introduction to the webgpu api},
  author={Kenwright, Benjamin},
  booktitle={Acm siggraph 2022 courses},
  pages={1--184},
  year={2022}
}

@misc{farshidi2023understandinguserintentmodeling,
      title={Understanding User Intent Modeling for Conversational Recommender Systems: A Systematic Literature Review}, 
      author={Siamak Farshidi and Kiyan Rezaee and Sara Mazaheri and Amir Hossein Rahimi and Ali Dadashzadeh and Morteza Ziabakhsh and Sadegh Eskandari and Slinger Jansen},
      year={2023},
      eprint={2308.08496},
      archivePrefix={arXiv},
      primaryClass={cs.IR},
      url={https://arxiv.org/abs/2308.08496}, 
}

@inproceedings{ZhangWGLM24,
  author       = {Erhan Zhang and
                  Xingzhu Wang and
                  Peiyuan Gong and
                  Yankai Lin and
                  Jiaxin Mao},
  editor       = {Grace Hui Yang and
                  Hongning Wang and
                  Sam Han and
                  Claudia Hauff and
                  Guido Zuccon and
                  Yi Zhang},
  title        = {USimAgent: Large Language Models for Simulating Search Users},
  booktitle    = {Proceedings of the 47th International {ACM} {SIGIR} Conference on
                  Research and Development in Information Retrieval, {SIGIR} 2024, Washington
                  DC, USA, July 14-18, 2024},
  pages        = {2687--2692},
  publisher    = {{ACM}},
  year         = {2024},
  url          = {https://doi.org/10.1145/3626772.3657963},
  doi          = {10.1145/3626772.3657963}
}

@misc{becker2024multiagentlargelanguagemodels,
      title={Multi-Agent Large Language Models for Conversational Task-Solving}, 
      author={Jonas Becker},
      year={2024},
      eprint={2410.22932},
      archivePrefix={arXiv},
      primaryClass={cs.CL},
      url={https://arxiv.org/abs/2410.22932}, 
}

@inproceedings{NarayananS08,
  author       = {Arvind Narayanan and
                  Vitaly Shmatikov},
  title        = {Robust De-anonymization of Large Sparse Datasets},
  booktitle    = {2008 {IEEE} Symposium on Security and Privacy {(SP} 2008), 18-21 May
                  2008, Oakland, California, {USA}},
  pages        = {111--125},
  publisher    = {{IEEE} Computer Society},
  year         = {2008},
  url          = {https://doi.org/10.1109/SP.2008.33},
  doi          = {10.1109/SP.2008.33}
}

@article{Ji_2023,
   title={Survey of Hallucination in Natural Language Generation},
   volume={55},
   ISSN={1557-7341},
   url={http://dx.doi.org/10.1145/3571730},
   DOI={10.1145/3571730},
   number={12},
   journal={ACM Computing Surveys},
   publisher={Association for Computing Machinery (ACM)},
   author={Ji, Ziwei and Lee, Nayeon and Frieske, Rita and Yu, Tiezheng and Su, Dan and Xu, Yan and Ishii, Etsuko and Bang, Ye Jin and Madotto, Andrea and Fung, Pascale},
   year={2023},
   month=mar, pages={1–38} 
}

@misc{wang2024anatomizingdeeplearninginference,
      title={Anatomizing Deep Learning Inference in Web Browsers}, 
      author={Qipeng Wang and Shiqi Jiang and Zhenpeng Chen and Xu Cao and Yuanchun Li and Aoyu Li and Yun Ma and Ting Cao and Xuanzhe Liu},
      year={2024},
      eprint={2402.05981},
      archivePrefix={arXiv},
      primaryClass={cs.LG},
      url={https://arxiv.org/abs/2402.05981}, 
}

@article{Zhao_2024,
   title={E-commerce Webpage Recommendation Scheme Base on Semantic Mining and Neural Networks},
   volume={4},
   ISSN={2790-1505},
   url={http://dx.doi.org/10.53469/jtpes.2024.04(03).20},
   DOI={10.53469/jtpes.2024.04(03).20},
   number={03},
   journal={Journal of Theory and Practice of Engineering Science},
   publisher={Century Science Publishing Co},
   author={Zhao, Wenchao and Liu, Xiaoyi and Xu, Ruilin and Xiao, Lingxi and Li, Muqing},
   year={2024},
   month=mar, pages={207–215} 
}

@techreport{W3C_IndexedDB_2015,
  author       = {Web Applications Working Group},
  title        = {Indexed Database API},
  institution  = {World Wide Web Consortium (W3C)},
  type         = {W3C Recommendation},
  number       = {REC-IndexedDB-20150108},
  year         = {2015},
  month        = {Jan},
  url          = {https://www.w3.org/TR/2015/REC-IndexedDB-20150108/}
}

@misc{AOLQueryLog2006,
  author       = {Pass, Gilad and Chowdhury, Abdur and Torgeson, Chris},
  title        = {AOL Query Log: 20 Million Queries from 650 000 Users (March–May 2006)},
  howpublished = {Web research dataset released by AOL Research},
  year         = {2006},
  note         = {Distributed as “AOL User Session Collection 500K”; anonymised; for research use only},
  url          = {https://www.kaggle.com/datasets/dineshydv/aol-user-session-collection-500k},
}

@misc{agarwal2021lookupexploratorysearchintent,
      title={Lookup or Exploratory: What is Your Search Intent?}, 
      author={Manoj K. Agarwal and Tezan Sahu},
      year={2021},
      eprint={2110.04640},
      archivePrefix={arXiv},
      primaryClass={cs.IR},
      url={https://arxiv.org/abs/2110.04640}, 
}

@misc{li2024bmlpbehaviorawaremlpheterogeneous,
      title={BMLP: Behavior-aware MLP for Heterogeneous Sequential Recommendation}, 
      author={Weixin Li and Yuhao Wu and Yang Liu and Weike Pan and Zhong Ming},
      year={2024},
      eprint={2402.12733},
      archivePrefix={arXiv},
      primaryClass={cs.IR},
      url={https://arxiv.org/abs/2402.12733}, 
}

@article{Williams1992,
  author    = {Ronald J. Williams},
  title     = {Simple Statistical Gradient-Following Algorithms for Connectionist Reinforcement Learning},
  journal   = {Machine Learning},
  volume    = {8},
  number    = {3--4},
  pages     = {229--256},
  year      = {1992},
  publisher = {Springer},
  doi       = {10.1007/BF00992696}
}

@techreport{W3C_WebWorkers_20210128,
  author       = {Web Applications Working Group},
  title        = {Web Workers},
  institution  = {World Wide Web Consortium (W3C)},
  type         = {W3C Working Group Note},
  number       = {NOTE-workers-20210128},
  year         = {2021},
  month        = {Jan},
  url          = {https://www.w3.org/TR/2021/NOTE-workers-20210128/}
}

@inproceedings{LuL0YLLLZLX25,
  author       = {Zhenyan Lu and
                  Xiang Li and
                  Dongqi Cai and
                  Rongjie Yi and
                  Fangming Liu and
                  Wei Liu and
                  Jian Luan and
                  Xiwen Zhang and
                  Nicholas D. Lane and
                  Mengwei Xu},
  editor       = {Wanxiang Che and
                  Joyce Nabende and
                  Ekaterina Shutova and
                  Mohammad Taher Pilehvar},
  title        = {Demystifying Small Language Models for Edge Deployment},
  booktitle    = {Proceedings of the 63rd Annual Meeting of the Association for Computational
                  Linguistics (Volume 1: Long Papers), {ACL} 2025, Vienna, Austria,
                  July 27 - August 1, 2025},
  pages        = {14747--14764},
  publisher    = {Association for Computational Linguistics},
  year         = {2025},
  url          = {https://aclanthology.org/2025.acl-long.718/}
}

@incollection{Brooke1996SUS,
  author    = {John Brooke},
  title     = {SUS: A ``Quick and Dirty'' Usability Scale},
  booktitle = {Usability Evaluation in Industry},
  editor    = {P. W. Jordan and B. Thomas and I. L. McClelland and B. Weerdmeester},
  pages     = {189--194},
  publisher = {Taylor \& Francis},
  year      = {1996}
}

\end{document}